# Forecasting Corn Yield With Machine Learning Ensembles

Mohsen Shahhosseini[1], Guiping Hu[1*] and Sotirios V. Archontoulis[2]

[1] *Department of Industrial and Manufacturing Systems Engineering, Iowa State University, Ames, IA, United States,*
[2] *Department of Agronomy, Iowa State University, Ames, IA, United States*

The emergence of new technologies to synthesize and analyze big data with high-performance computing has increased our capacity to more accurately predict crop yields. Recent research has shown that machine learning (ML) can provide reasonable predictions faster and with higher flexibility compared to simulation crop modeling. However, a single machine learning model can be outperformed by a "committee" of models (machine learning ensembles) that can reduce prediction bias, variance, or both and is able to better capture the underlying distribution of the data. Yet, there are many aspects to be investigated with regard to prediction accuracy, time of the prediction, and scale. The earlier the prediction during the growing season the better, but this has not been thoroughly investigated as previous studies considered all data available to predict yields. This paper provides a machine leaning based framework to forecast corn yields in three US Corn Belt states (Illinois, Indiana, and Iowa) considering complete and partial in-season weather knowledge. Several ensemble models are designed using blocked sequential procedure to generate out-of-bag predictions. The forecasts are made in county-level scale and aggregated for agricultural district and state level scales. Results show that the proposed optimized weighted ensemble and the average ensemble are the most precise models with RRMSE of 9.5%. Stacked LASSO makes the least biased predictions (MBE of 53 kg/ha), while other ensemble models also outperformed the base learners in terms of bias. On the contrary, although random k-fold cross-validation is replaced by blocked sequential procedure, it is shown that stacked ensembles perform not as good as weighted ensemble models for time series data sets as they require the data to be non-IID to perform favorably. Comparing our proposed model forecasts with the literature demonstrates the acceptable performance of forecasts made by our proposed ensemble model. Results from the scenario of having partial in-season weather knowledge reveals that decent yield forecasts with RRMSE of 9.2% can be made as early as June 1st. Moreover, it was shown that the proposed model performed better than individual models and benchmark ensembles at agricultural district and state-level scales as well as county-level scale. To find the marginal effect of each input feature on the forecasts made by the proposed ensemble model, a methodology is suggested that is the basis for finding feature importance for the ensemble model. The findings suggest that weather features corresponding to weather in weeks 18–24 (May 1st to June 1st) are the most important input features.

# INTRODUCTION

Providing 11% of total US employment, agriculture and its related industries are considered as a significant contributor to the US economy, with $1.053 trillion of US gross domestic product (GDP) in 2017 (USDA Economic Research Center, 2019). Crop yield prediction is of high significance since it can provide insights and information for improving crop management, economic trading, food production monitoring, and global food security. In the past, farmers relied on their experiences and past historical data to predict crop yield and make important cropping decisions based on the prediction. However, the emergence of new technologies, such as simulation crop models and machine learning in the recent years, and the ability to analyze big data with high-performance computing has resulted in more accurate yield predictions (Drummond et al., 2003; Vincenzi et al., 2011; González Sánchez et al., 2014; Jeong et al., 2016; Pantazi et al., 2016; Cai et al., 2017; Chlingaryan et al., 2018; Crane-Droesch, 2018; Basso and Liu, 2019; Shahhosseini et al., 2019c).

Forecasting crop production is different from prediction as it requires interpreting future observations only using the past data (Griffiths et al., 2010; Johnson, 2014; Brockwell and Davis, 2016; Cai et al., 2017). Previous studies considered all the data for forecasting, while the next challenge is to consider partial data as it reflects reality better if we are to use a forecast model to inform farmers and decision makers. Also, the scale of prediction is of interest. Yet we do not know if predictions are more accurate at a finer (county) or course (agricultural district) scale. Previous research by Sakamoto et al. (2014) and Peng et al. (2018) suggested better prediction accuracy for course scale compared to a finer scale.

Simulation crop modeling has a reasonable prediction accuracy, but due to user skill, data calibration requirements, long runtimes, and data storage constraints, it is not as easily applicable as machine learning (ML) models (Drummond et al., 2003; Puntel et al., 2016; Shahhosseini et al., 2019c). On the other hand, ML has enjoyed a wide range of applications in various problems, including ecological predictive modeling, because of its ability in dealing with linear and nonlinear relationships, non-normal data, and quality of results along with significantly lower runtimes (De'ath and Fabricius, 2000).

Generally, supervised learning is categorized into regression and classification problems, based on the type of response variables. Many studies have approached regression problems, in which the response variable is continuous, with machine learning to solve an ecological problem (James et al., 2013). These studies include but not limited to crop yield predictions (Drummond et al., 2003; Vincenzi et al., 2011; González Sánchez et al., 2014; Jeong et al., 2016; Pantazi et al., 2016; Cai et al., 2017; Chlingaryan et al., 2018; Crane-Droesch, 2018; Basso and Liu, 2019; Khaki and Wang, 2019; Shahhosseini et al., 2019c; Emirhüseyinoğlu and Ryan, 2020; Khaki et al., 2020), crop quality (Hoogenboom et al., 2004; Karimi et al., 2008; Mutanga et al., 2012; Shekoofa et al., 2014; Qin et al., 2018; Ansarifar and Wang, 2019; Khaki et al., 2019; Lawes et al., 2019; Moeinizade et al., 2019), water management (Mohammadi et al., 2015; Feng et al., 2017; Mehdizadeh et al., 2017), soil management (Johann et al., 2016; Morellos et al., 2016; Nahvi et al., 2016), and others.

Studies show that a single machine learning model can be outperformed by a "committee" of individual models, which is called a machine learning ensemble (Zhang and Ma, 2012). Ensemble learning is proved to be effective as it can reduce bias, variance, or both and is able to better capture the underlying distribution of the data in order to make better predictions, if the base learners are diverse enough (Dietterich, 2000; Pham and Olafsson, 2019a; Pham and Olafsson, 2019b; Shahhosseini et al., 2019a; Shahhosseini et al., 2019b). The usage of ensemble learning in ecological problems is becoming more widespread; for instance, bagging and specifically random forest (Vincenzi et al., 2011; Mutanga et al., 2012; Fukuda et al., 2013; Jeong et al., 2016), boosting (De'ath, 2007; Heremans et al., 2015; Belayneh et al., 2016; Stas et al., 2016; Sajedi-Hosseini et al., 2018), and stacking (Conţiu and Groza, 2016; Cai et al., 2017; Shahhosseini et al., 2019a), are some of the ensemble learning applications in agriculture. Although, there have been studies using some of the ensemble methods in the agriculture domain, to the best of our knowledge, there is no study to compare the effectiveness of ensemble learning for ecological problems, especially when there are temporal and spatial correlations in the data.

In this paper, we develop machine learning algorithms to forecast corn yields in three US Corn Belt states (Illinois, Indiana, and Iowa) using data from 2000 to 2018. These three states together produce nearly 50% of the total corn produced in the USA, which has an economic value of $20 billion per year (NASS, 2019). In 2019, corn was the largest produced crop in the United States (Capehart and Proper, 2019) and with the increasing movement towards ethanol to replace gas in cars, it is almost necessary to increase the amount of corn being produced. Hence, forecasting the corn yield for important US corn producing states could provide valuable insights for decision making.

Therefore, we design several ML and ML ensemble models using blocked sequential procedure (Cerqueira et al., 2017; Oliveira et al., 2019) to generate out-of-bag predictions and evaluate their performance when forecasting corn yields. In addition, we investigate the effect of having complete or partial in-season weather knowledge when forecasting yields. The forecasts are made in three scales: county, agricultural district, and state level, and the state-level forecasts are compared with USDA NASS forecasts. Furthermore, a methodology to calculate partial dependency of the proposed ensemble model is proposed which can quantify the marginal effect of changing each input feature on the forecasts made be the ML ensemble model. Based on the computed partial dependencies, a measure to calculate the importance of input features from optimized weighted ensemble model is proposed which ranks input features based on the variations in their partial dependency plots (PDPs). This analysis can help prioritize which data to be collected in the future and inform agronomists to explain causes of high or low yield levels in some years.

The remainder of this paper is organized as follows. The data and methodologies are described in *Materials and Methods*. *Results and Discussion* is dedicated to the model performance results, discussions, and potential improvements. Finally, the paper concludes with the findings in the *Conclusion*.

## MATERIALS AND METHODS

The designed machine learning models aim at forecasting corn yield in three US Corn Belt states with a data set including environmental (soil and weather) and management variables for two different scenarios; complete knowledge of in-season weather, partial knowledge of in-season weather (until August 1st) and three scales; county, agricultural district, and state level. We selected three major corn production states in the US Corn Belt to explore our research questions considering also the computational complexity of the developed ensemble models.

The data inputs used to drive ML were approximately the same as those that were used to drive a crop model predictions (APSIM) in this region (Archontoulis and Licht, 2019). They were selected because all of them are agronomically relevant for yield predictions (Archontoulis et al., 2019). The data contains several soil parameters at a 5 km resolution (Soil Survey Staff et al., 2019), weather data at 1 km resolution (Thornton et al., 2012), crop yield data at different scales (NASS, 2019), and management information at the state level (NASS, 2019).

### Data Set
County-level historically observed corn yields were obtained from the USDA National Agricultural Statistics Service (NASS, 2019) for years 2000–2018. A data set was developed containing observed information of corn yields, management (plant population and planting date), and environment (weather and soil) features.

- *Plant population*: plant population measured in plants/acre, downloaded from USDA NASS
- *Planting progress (planting date)*: The weekly cumulative percentage of corn planted over time within each state (NASS, 2019)
- *Weather*: 7 weather features aggregated weekly, downloaded from Daymet (Thornton et al., 2012)
- Daily minimum air temperature in degrees Celsius.
- Daily maximum air temperature in degrees Celsius.
- Daily total precipitation in millimeters per day

- Shortwave radiation in watts per square meter
- Water vapor pressure in pascals
- Snow water equivalent in kilograms per square meter
- Day length in seconds per day
- *Soil*: The following soil features were considered in this study: soil organic matter, sand content, clay content, soil pH, soil bulk density, wilting point, field capacity, saturation point, and hydraulic conductivity. Because these features change across the soil profile, we used different values for different soil layers, which resulted in 180 features for soil characteristics of the selected locations, downloaded from the Web Soil Survey (Soil Survey Staff et al., 2019)
- *Yield*: Annual corn yield data, downloaded from the USDA National Agricultural Statistics Service (NASS, 2019)

The developed data set consists of 5,342 observations of annual average corn yields for 293 counties across three states on the Corn Belt and 597 input features mentioned above. The reason to choose these components as the explanatory features is that the factors affecting yield performance are mainly environment, genotype, and management. Weather and soil features were included in the data set to account for environment component, as well as management, but since there is no publicly available genotype data set, the effect of genotype on the yield performance is not considered. In this study we used many input parameters that are probably less likely to be available in other parts of the world. In this case we recommend use of gridded public soil or weather databases used to drive global crop production models (Rosenzweig et al., 2013; Hengl et al., 2014; Elliott et al., 2015; Han et al., 2019).

### Data Pre-Processing
Data pre-processing tasks were performed before training the machine learning models. First off, the data of the years 2016–2018 were reserved as the test subset and the remaining data were used to build the models. Second, all input variables were scaled and transformed to a range between 0 and 1 to prevent the magnitude of some features from misleading the machine learning models. Third, new features were constructed that account for the yearly trends in the yields, and finally, random forest-based feature selection was performed to avoid overfitting in model training.

#### Feature Construction for the Yearly Trends
**Figures 1A, B** suggest an increasing trend in the corn yields for the locations under study. This trend is due to improved genetics (cultivars), improved management, and other technological advances such as farming equipment (range of yield increase was from 32 to 189 kg/ha/year). Since there is no feature in the input variables that can explain this observed trend, we decided to add new features to the developed data set that can explain the trend

Temperature is one of the many factors that influence historical yield increase. Other factors are changes in weather (precipitation), increase in plant density, improved genetics, improved planting technology, and improvements in soil and crop management over time. Because there is not enough information to separate the contribution of each factor with the available data, we simply considered all these factors as one factor in this study.

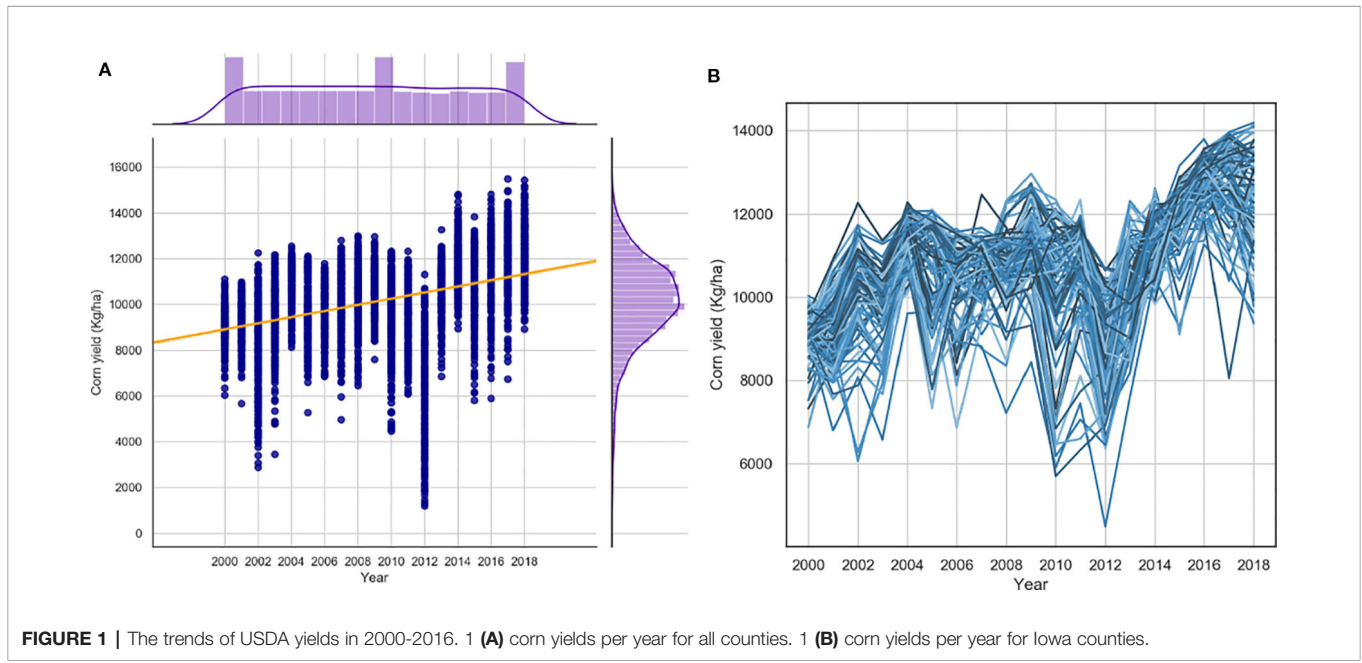

FIGURE 1 | The trends of USDA yields in 2000-2016. 1 (A) corn yields per year for all counties. 1 (B) corn yields per year for Iowa counties.

Two measures were done to account for the trend in yields.

1) To observe the trend in corn yields, a new feature (yield_trend) was created. A linear regression model was built for each location as the trends for each site tend to be different. The independent and dependent variables of this linear regression model were comprised of the year (YEAR) and yield (Y), respectively. Afterwards, the predicted value for each data point ($\hat{Y}$) is added as the value of the new feature. The data used for fitting this linear trend model was only training data, and for finding the corresponding values of the newly added feature for the test set observations, the prediction made by this trend model for the data of the test years ($\hat{Y}_{i,test} = b_{0_i} + b_{1_i} YEAR_{i,test}$) was used. The trend value ($\hat{Y}_i$) calculated for each location (i) that is added to the data set as a new feature is shown in the following equation.

$$\hat{Y}_i = b_{0_i} + b_{1_i} YEAR_i \quad (1)$$

2) Moreover, another new variable (yield_avg) was constructed that defines the average yield of each year for each state when considering the training data. The procedure to find the average value of the yields of each state (j) as the values of the new feature is shown mathematically in the equation (2).

$$yield\_avg_j = average(yield_j) \quad (2)$$

It should be noted that the corresponding values of this feature for the unseen test observations are calculated as follows. The last training year (2015) in each state is used as a baseline, and the average increment in the average yield of each state is used as a measure of increase in the state-wide average yield. The following equation demonstrates the calculation of the values of newly created feature for unseen test observations of state j (years 2016–2018).

$$yield_{avg_{j,t}} = average(yield_{j,2015}) \left[ 1 + average\left( \frac{yield_{avg_{j,n}} - yield_{avg_{j,n-1}}}{yield_{avg_{j,n-1}}} \right) \right]^{t-2015} \quad (3)$$

where j shows each state, t denotes the test year (2016–2018), and n represents the training year starting from the year 2001.

### Three-Stage Feature Selection

As mentioned earlier the developed data set has a small observation-to-feature ratio (5,342/597), which may lead to overfitting on the training data because of its sparsity and large number of input variables, and the built models may not generalize well to the unseen observations. To address this problem, we conduct a three-stage feature selection procedure to select only the best input variables to include in our model and reduce the data set dimensions. To this end, first, a feature selection based on expert knowledge was performed. Weather features for the period after harvesting and before planting were removed. In addition, the cumulative planting progress features for the weeks before planting were removed since they didn't include any information. This reduced the number of independent variables from 597 to 383. In the second stage, a permutation importance feature selection procedure based on random forest learning algorithm was conducted. Specifically, the 80 most important input features ranked by permutation importance of random forest model built on the training set were included in the training data set. The final stage of feature selection was a filter-based feature selection based on Pearson correlation values. In this procedure, assuming linear relationships between independent variables, features that were highly correlated (with a Pearson correlation higher than 0.9) were identified and from each pair of linearly dependent features only one feature were remained in the data set. This can be justified by the fact that when two features are highly correlated, they have almost the same effect on the response variable; hence one of them is redundant. This three-stage process is depicted in the **Figure 2**. It should be noted that the constructed features for yearly yield trends were kept in the analysis data set.

## Hyperparameter Tuning and Model Selection
### Walk-Forward Cross-Validation

Optimizing hyperparameters of machine learning models could improve the prediction accuracy and generalizability of the trained models. Traditionally, k-fold cross-validation is used to find the best hyperparameter values using only training data. However, the assumption of the data being independent and identically distributed (IID) does not hold for time series data sets and disregarding this assumption will result in a cross-validation scheme that does not emulate the test distribution well (Bergmeir et al., 2018). Hyndman and Athanasopoulos (2018) introduced a walk-forward cross-validation procedure for time series analysis. In this method, a set of validation sets is defined, each consisting of data from a single point in time. The training set is formed by all the time points that occurred before each validation observation. Therefore, future observations are not used in forecasting. Hence, to optimize the hyperparameter values of machine learning models and select the best models only using the training set, a variation of the walk-forward cross-validation introduced in Hyndman and Athanasopoulos (2018) is used, where the training part of each fold is assumed to have the same size. This assumption was made aiming at reducing the computational time after observing the prediction results when using walk-forward cross-validation procedure proposed by Hyndman and Athanasopoulos (2018). In each fold, the training set size is assumed to be 8 years, and the following year is considered as validation set.

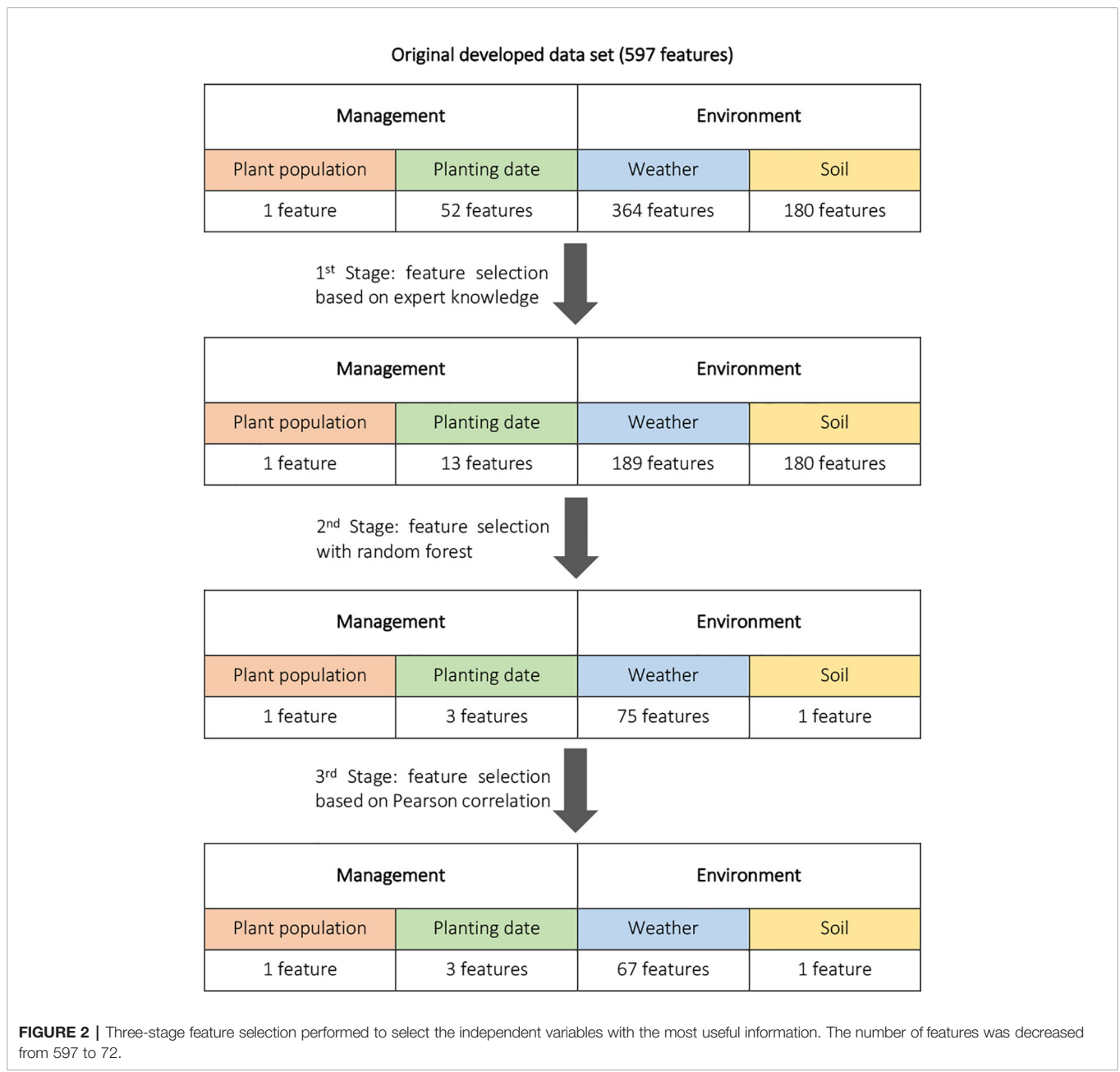

**FIGURE 2 |** Three-stage feature selection performed to select the independent variables with the most useful information. The number of features was decreased from 597 to 72.

## Bayesian Search

Assuming an unknown underlying distribution, Bayesian optimization intends to approximate the unknown function with surrogate models such as Gaussian process. Bayesian optimization is mainly different from other search methods in incorporating prior belief about the underlying function and updating it with new observations. This difference makes Bayesian search for hyperparameter tuning faster than exhaustive grid search while finding a better solution compared to random search. Bayesian optimization collects instances with the highest information in each iteration by making a balance between exploration (exploring uncertain hyperparameters) and exploitation (gathering observations from hyperparameters close to the optimum) (Snoek et al., 2012). Thus, Bayesian search was selected as the hyperparameter tuning search method under the look-forward cross-validated procedure. Bayesian optimization is conducted with the objective of minimizing training mean squared error (MSE) on the search space consisting of hyperparameter values and using Tree-structured Parzen Estimator Approach (TPE) which uses the Bayes rule to construct the surrogate model (Bergstra et al., 2011).

## Analyzed Models

Well-performing ensemble models require the base learners to exhibit a certain element of "diversity" in their predictions along with retaining good performance individually (Brown, 2017). Therefore, a set of different models was selected and trained including linear regression, LASSO regression, Extreme Gradient Boosting (XGBoost), LightGBM, and random forest. Random forest uses ensembles of fully-grown trees, and therefore tend to have lower bias and higher variance. Differently, gradient boosting is iteratively built on weak learners that tend to be on the opposite end of the bias/variance tradeoff. Linear regression is also added as a benchmark, and LASSO regression is included due to its intrinsic feature selection. In addition, multiple two-level stacking ensemble models, as well as average ensemble and exponentially weighted average ensemble (EWA) were constructed and evaluated on test unseen observations. Furthermore, an optimized weighted ensemble model that accounts for both bias and variance of the predictions was proposed that can use out-of-bag predictions to find the optimal weights in making optimal weighted ensembles. The mentioned models can deal with features that have linear or nonlinear correlation with the response variable.

### Linear Regression

Assuming a linear relationship between the predictors and the response variable, normal distribution of residuals (normality), absence of correlation between predictors (no multicollinearity), and similar variance of error across predictors (homoscedasticity), linear regression predicts a quantitative response based on multiple predictor variables. A multiple linear regression model is in the following form (James et al., 2013).

$$Y = \beta_0 + \beta_1 X_1 + \beta_2 X_2 + \ldots + \beta_p X_p + \epsilon \quad (4)$$

in which $Y$ is the response variable, $X_j$ are the independent variables, $\beta_j$ are the coefficients, and $\epsilon$ is the error term. The coefficients are estimated by minimizing the loss function $L$, as shown below.

$$L = \sum_{i=1}^{n}(y_i - \hat{y}_i)^2$$
$$= \sum_{i=1}^{n}\left(y_i - \hat{\beta}_0 - \hat{\beta}_1 X_{i1} - \hat{\beta}_2 X_{i2} - \ldots - \hat{\beta}_p X_{ip}\right)^2 \quad (5)$$

where $\hat{y}_i$ is the prediction for $y_i$.

### LASSO Regression

Least absolute shrinkage and selection operator (LASSO) is a regularization method that is able to exclude some of the variables by setting their coefficient to zero (James et al., 2013). A penalty term ($|\beta_i|$) is added to the linear regression model in LASSO which is able to shrink coefficients towards zero (L1 regularization). The loss function of LASSO is as follows (Tibshirani, 1996).

$$L = \sum_{i=1}^{n}(y_i - \hat{y}_i)^2 + \lambda \sum_{j=1}^{p} |\beta_j| \quad (6)$$

where $\lambda$ is the shrinkage parameter that needs to be determined before performing the learning task.

### XGBoost and LightGBM

Gradient boosting, a tree-based ensemble method, makes predictions by sequentially combining weak prediction models. In other words, gradient boosting predicts by learning from mistakes made by previous predictors. In this study, we made use of two relatively new and fast implementations of gradient boosting: XGBoost and LightGBM. XGBoost, proposed in 2016, is capable of handling sparse data and makes use of an approximation algorithm, Weighted Quantile Sketch, to determine splits and speed-up the learning process (Chen and Guestrin, 2016). LightGBM from Microsoft, published in 2017, introduced two ideas to improve performance and reduce the computational time. First, gradient-based one-side sampling helps select the most informative observations. Second, Exclusive Feature Bundling (EFB) takes advantage of data sparsity and bins similar input features (Ke et al., 2017).

### Random Forest

Bootstrap aggregating (Bagging) is another tree-based ensemble model, which tries to reduce the variance of predictions, and consequently increases the model's generalizability by generating multiple trees from training data using sampling with replacement (Breiman, 1996). Random forest is a special case of bagging ensemble in which each tree depends on a random value, number of predictors chosen as split candidates in each iteration. (Breiman, 2001). This makes random forest superior than bagging since random forest de-correlates the trees. In addition, random forest makes use of observations not included in the bootstrapped samples (out-of-bag observations) to compute error rates (Cutler et al., 2007).

## Stacked Generalization

Stacked generalization aims to minimize the generalization error of some ML models by performing at least one more level of learning task using the outputs of ML base models as inputs and the actual response values of some part of the data set (training data) as outputs (Wolpert, 1992). Stacked generalization assumes the data to be IID and performs a $k$-fold cross-validation to generate out-of-bag predictions for validation set of each fold. Collectively, the $k$ out-of-bag predictions create a new training set for the second level learning task, with the same size of the original training set (Cai et al., 2017). However, here the IID assumption of the data does not hold, and we cannot use $k$-fold cross-validation to generate out-of-bag predictions. To work around this issue, blocked sequential procedure (Cerqueira et al., 2017; Oliveira et al., 2019) was used to generate inputs of the stacked generalization method only using past data (see **Figure 3**).

The following steps describe this procedure:

a. Consider first 8 years as training and the following year as validation set.
b. Train each base learner on the training data and make predictions for the validation set (out-of-bag predictions).
c. Record the out-of-bag predictions and move the training and validation sets one year forward.
d. Repeat (a)–(c) until reach the end of original training set.

Here it should be noted that the size of the generated out-of-bag predictions matrix is smaller than the original training set since it does not include first 8 years of data in the validation sets.

As the second level predictive model, four machine learning models were selected resulting in four stacked generalization models:

1. Stacked regression: linear regression as the second level model
2. Stacked LASSO: LASSO regression as the second level model
3. Stacked random forest: random forest as the second level model
4. Stacked LightGBM: LightGBM as the second level model

## Proposed Optimized Weighted Ensemble

Optimized weighted ensembles can be created with an optimization model. Due to the tradeoff between bias and variance of the prediction, the optimized ensemble should be able to predict with the least possible bias and variance. Specifically, we take advantage of bias and variance decomposition as follows.

$$E\left[\left(f(x) - \hat{f}(x)\right)^2\right]$$
$$= \left(Bias\left[\hat{f}(x)\right]\right)^2 + Var\left[\hat{f}(x)\right] + Var(\epsilon) \quad (7)$$

Based on bias and variance tradeoff, the objective function of the optimization problem can be mean squared error (MSE) of out-of-bag predictions for the ensemble (Hastie et al., 2009). The out-of-bag predictions matrix created previously can be used as an emulator of unseen test observations (Shahhosseini et al., 2019b). Using the out-of-bag predictions, we propose an optimization problem which is a nonlinear convex optimization problem as follows.

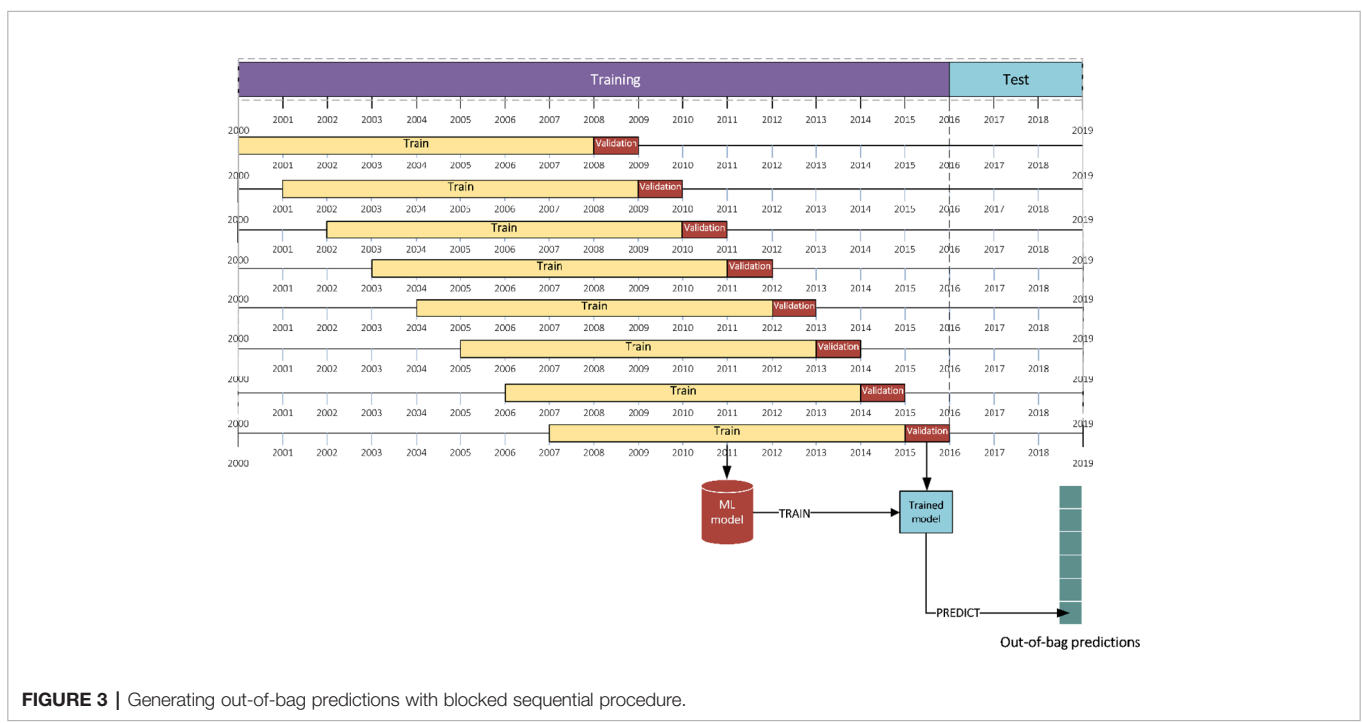

**FIGURE 3 |** Generating out-of-bag predictions with blocked sequential procedure.

$$Min \frac{1}{n}\sum_{i=1}^{n}(y_i - \sum_{j=1}^{k} w_j \hat{y}_{ij})^2$$
$$s.t.$$
$$\sum_{j=1}^{k} w_j = 1$$
$$w_j \geq 0, \quad \forall j = 1,...,k.$$
(8)

where $w_j$ is the weights corresponding to base model $j$ ($j = 1,..., k$), $n$ is the total number of instances ($n$ is smaller than the number of original training set observations because the first 8 years of training data never were included in the validation set), $y_i$ is the true value of observation $i$, and $\hat{y}_{ij}$ is the prediction of observation $i$ by base model $j$. Since other ensemble learning models such as stacking strictly require the data to be IID, and that the proposed model does not have such requirement, we expect this model to outperform the stacking ensembles as well as base models.

### Average Ensemble

Average ensemble is the weighted average of out-of-bag predictions made by base learners when they all have equal weights ($w_j = 1/k$). When the base machine learning models are diverse enough, the average ensemble can perform better than each of base learners (Brown, 2017).

### Exponentially Weighted Average Ensemble (EWA)

Exponentially weighted average ensemble is different from other ensemble creation methods, as it does not require the out-of-bag predictions but the out-of-bag errors. In fact, the weights for each model can be computed using its past performance. In this case, we find the prediction error of out-of-bag predictions made by each ML base learner and calculate their corresponding weights as follows (Cesa-Bianchi and Lugosi, 2006).

$$w_j = \frac{exp(-e_j)}{\sum_{j=1}^{k} exp(-e_j)}$$
(9)

where $e_j$ is the out-of-bag prediction error of base learner $j$.

## Statistical Performance Metrics
### Root Mean Squared Error (RMSE)

Root mean squared error (RMSE) is defined as the square root of the average squared deviation of predictions from actual values (Zheng, 2015).

$$RMSE = \sqrt{\frac{\sum_i (y_i - \hat{y}_i)^2}{n}}$$
(10)

where $y_i$ denotes the actual values, $\hat{y}_i$ is the predictions and $n$ denotes the number of data points.

### Relative Root Mean Squared Error (RRMSE)

Relative root mean squared error (or normalized root mean squared error) is the RMSE normalized by the mean of the actual values and is often expressed as percentage. Lower values for RRMSE are preferred.

$$RRMSE = \frac{RMSE}{\bar{y}}$$
(11)

### Mean Bias Error (MBE)

Mean bias error (MBE) is a measure to describe the average bias in the prediction.

$$MBE = \frac{\sum_i (\hat{y}_i - y_i)}{n}$$
(12)

### Mean Directional Accuracy (MDA)

Mean directional accuracy (MDA) provides a metric to find the probability that the prediction model can detect the correct direction of time series (Cicarelli, 1982; Schnader and Stekler, 1990). While other metrics such as RMSE, RRMSE, and MBE are crucial to evaluate the performance of the forecast, the directional movement of the forecast is important to understand the capture of trend. This measure is commonly used in economics and macroeconomics studies.

$$MDA = \frac{\sum_t 1_{sign(y_t - y_{t-1}) == sign(\hat{y}_t - y_{t-1})}}{n}$$
(13)

where $y_t$ and $\hat{y}_t$ are actual values and prediction at time $t$, $\mathbf{1}$ is the indicator function, and $sign(\cdot)$ denotes the sign function.

## RESULTS AND DISCUSSION

After presenting the numerical results of designed forecasting ML models and comparing them with the literature, this section discusses the effect of in-season weather information on the quality of forecasts by comparing the prediction accuracy of designed ensemble models on different subsets of in-season weather information. In addition, we propose an approach to calculate the partial dependency of the input features to the forecasts made by the optimized weighted ensemble model and interpret the subsequent partial dependence plots. Moreover, a method for computing importance of input features based on partial dependency is designed and implemented to find the most influential independent variables for optimized weighted ensemble.

### Numerical Results

The designed machine learning models were evaluated on two different scenarios: complete knowledge of in-season weather and partial knowledge of in-season weather (discussed in *Partial Knowledge of In-Season Weather Information*). In addition, the results were aggregated in different scales of county, agricultural district, and state levels. The models are run on a computer equipped with a 2.6 GHz Intel E5-2640 v3 CPU, and 128 GB of RAM (see **Table 1** for computational times).

**Table 2** summarizes the performance of ML models considering complete in-season weather knowledge on county-level scale.

As **Table 2** shows, from the base ML models, random forest makes the least prediction error based on RMSE and RRMSE indices. The MBE results show that the linear regression and LASSO regression are the only prediction models that overestimate the true values, and other ML models underestimate the yields. Furthermore, random forest predictions are not as biased as other base learners based on MBE values.

Ensemble models provide better performance compared to the base learners. The proposed optimized weighted ensemble and the average ensemble are the most precise models with RRMSE of 9.5%, which improves the prediction error of best base learner (random forest) by about 8%. Stacked LASSO makes the least biased predictions (MBE of 53 kg/ha), while other ensemble models also outperformed the base learners in terms of bias (see **Figure 4**).

It can be seen that weighted ensembles (optimized weighted ensemble, average ensemble, and exponentially weighted ensemble) outperform base learners and stacked ensembles. This can be explained by the IID requirement of stacking models. Although random k-fold cross-validation was replaced by blocked sequential procedure to generate out-of-bag predictions, it seems that stacked ensemble models will not perform as good as weighted ensemble models for non-IID data sets. Regarding mean directional accuracy (MDA) of year 2018 based on year 2017, Stacked regression predicted the correct direction of corn yields 60% of the time, while optimized weighted ensemble model predictions are on the right direction 57% of the time.

Evaluating the performance of designed ML models when predicting test observations from different years suggests that weighted ensemble models are more accurate than other models for years 2016–2018 (see **Figure 5**). Furthermore, almost all models predicted the data from year 2017 with the least error and the data from year 2016 with the highest prediction error. **Figure 5** further proves that the weighted ensembles can take better advantage of diversity in the base learners than stacked ensembles.

The performance of our proposed optimized weighted ensemble model is also compared to the models developed in similar studies that tried to use machine learning to predict US corn yield. Jeong et al. (2016) could predict US corn yield with 30 years of data using random forest with the prediction RRMSE of 16.7%; while Crane-Droesch (2018) could achieve out-of-bag USDA corn prediction error of 13.4% using semiparametric neural network with a data set comprised of the information for years 1979–2016. Kim et al. (2019) designed a model which predicted cross-validation out-of-bag samples with a RRMSE of 7.9% (**Table 3**). It should be noted that because of non-IID nature of yield prediction data sets, it is not entirely appropriate to demonstrate cross-validation out-of-bag errors as the estimators of the true error. The presented error of our model is drawn from testing the developed model on unseen observations of future years.

Based on the results, the purpose of analysis can make one or more models more favorable against others. For instance, if the objective is to forecast corn yields with the lowest prediction error, weighted ensemble models should be selected; whereas, in the event that the goal is to detect the correct forecast direction, stacked LASSO regression could be chosen. However, the overall performance of weighted ensemble models, with having the least prediction error and acceptable bias and a quite high probability in detecting the right forecast direction, is better than other models.

**Table 4** summarizes the performance of the designed models when the forecasts are aggregated on agricultural district and state levels. Total area harvested was used as the measure to compute weighted average of county-level yields to obtain agricultural district and state-level corn yields. The results are in line with the county-level forecasts and optimized weighted ensemble and average ensemble as well as stacked LightGBM outpace base learners and other ensemble models in term of prediction error (RRMSE). Mean directional accuracy results are a bit different from county-level analysis and the reason seems to be smaller number of data points. Linear regression and LASSO appear to be

**TABLE 1** | Training and prediction times of designed ML models.

| ML Model | Training time (milliseconds) | Prediction time (milliseconds) |
| --- | --- | --- |
| Linear regression | 14 | 1.17 |
| LASSO | 9 | 1.19 |
| XGBoost | 5,973 | 6.58 |
| LightGBM | 2,229 | 36.84 |
| Random forest | 13,382 | 14.09 |
| Stacked regression | 91,558 | 0.50 |
| Stacked LASSO | 91,558 | 0.50 |
| Stacked random f. | 91,625 | 1.93 |
| Stacked LightGBM | 91,642 | 6.64 |
| Optimized w. ensemble* | 92,283 | 0.03 |
| Average ensemble | 91,556 | 0.03 |
| EWA | 92,300 | 0.03 |

The proposed model is distinguished with (*).

**TABLE 2** | Summary of designed county-level models performance.

| ML Model | RMSE (kg/ha) | RRMSE (%) | MBE (kg/ha) | MDA (%) (2018 – 2017) |
| --- | --- | --- | --- | --- |
| Linear regression | 1,533 | 12.87% | 599 | 50.79% |
| LASSO | 1,298 | 10.90% | 639 | 55.95% |
| XGBoost | 1,525 | 12.80% | −902 | 53.57% |
| LightGBM | 1,337 | 11.23% | −530 | 46.83% |
| Random forest | 1,242 | 10.43% | −387 | 55.16% |
| Stacked regression | 1,149 | 9.65% | 55 | 59.52% |
| Stacked LASSO | 1146 | 9.62% | 53 | 55.16% |
| Stacked random f. | 1,257 | 10.56% | −260 | 49.21% |
| Stacked LightGBM | 1,173 | 9.85% | −180 | 46.03% |
| Optimized w. ensemble* | 1,138 | 9.56% | 168 | 56.75% |
| Average ensemble | 1,137 | 9.54% | −116 | 56.75% |
| EWA | 1,148 | 9.64% | −149 | 56.35% |

The proposed model is distinguished with (*).

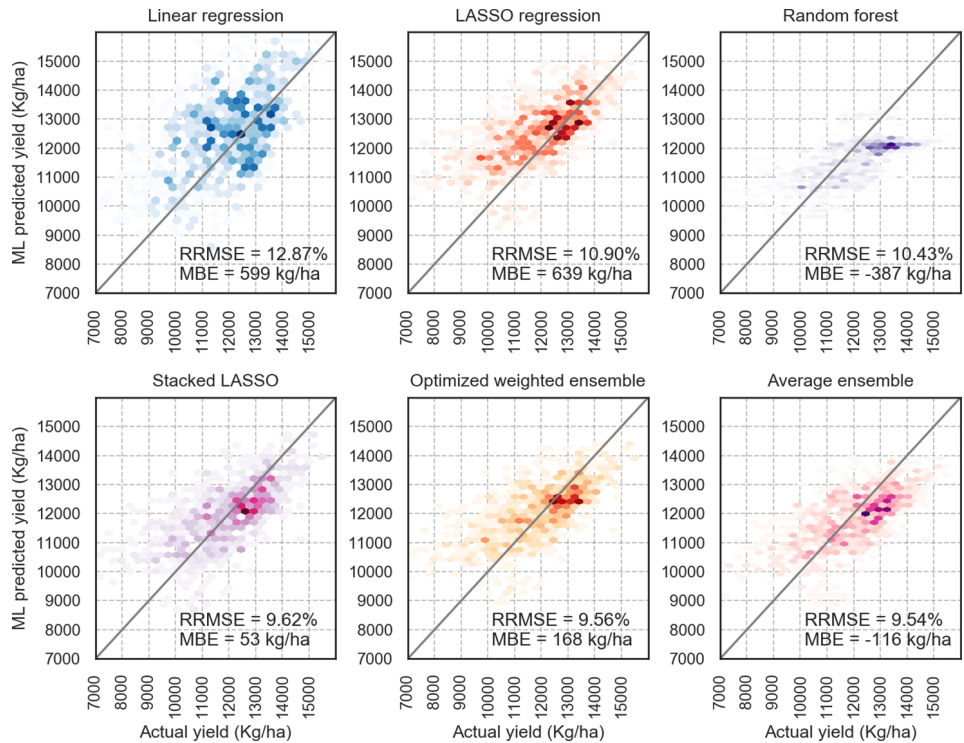

**FIGURE 4** | X–Y plots of some of the designed models; Optimized weighted ensemble and Average ensemble made predictions closer to the diagonal line; the color intensity shows the accumulation of the data points.

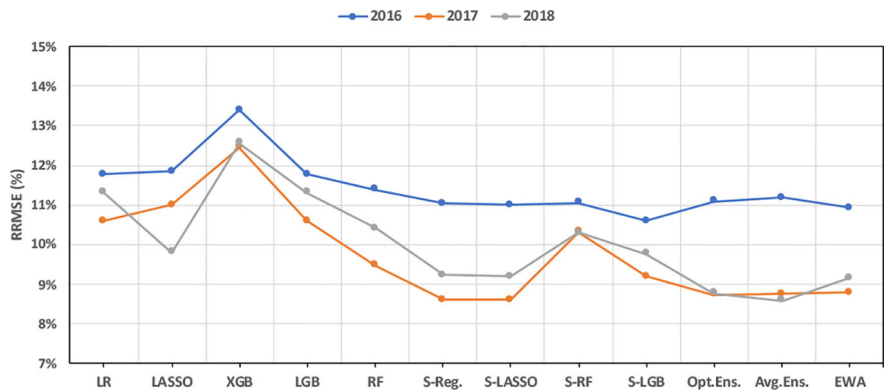

**FIGURE 5** | Performance of ML models in predicting test observations from different years.

the only base learners that overestimate the yields and have a higher probability to predict in the correct forecast direction.

## Partial Knowledge of In-Season Weather Information

To evaluate the impact of partial in-season weather knowledge on corn yield forecasts, the machine learning ensemble models were trained on a subset of weather features, including information from planting time up to June 1st, July 1st, August 1st, September 1st, and October 1st. Hyperparameter tuning and model selection have been done separately when considering each scenario. **Figure 6** demonstrates the RRMSE of ensemble forecasts when having partial in-season weather information. As the figure suggests, although the forecasts become more accurate with more recent weather data, decent forecasts can be made from weighted ensemble models as early as June 1st. This is a very important result because maize market price is usually high during that period (due to

TABLE 3 | Comparing prediction error of the proposed model (optimized weighted ensemble) with the literature. The error values of some studies were converted from different units to Kg/ha to have the same unit.

| | Data years | Forecast level | Forecast date | Test set | Developed model | RMSE (kg/ha) | RRMSE (%) |
|---|---|---|---|---|---|---|---|
| **Optimized w. ensemble*** | 2000–2018 | County | Oct | 2016-2018 | Optimal weighted ensemble | 1,138 | 9.5% |
| Bolton and Friedl (2013) | 2004–2008 | County | Sep | 2009 | MODIS[1]-based Linear regression | 809 | 8.0% |
| Johnson (2014) | 2006–2011 | County | Oct | 2012 | Cubist | 1,260 | 17.1% |
| Sakamoto et al. (2014) | 2008–2011 | State | Aug | 2002–2007 & 2012 | MODIS-based bias correction | 950 | 11.8% |
| Jeong et al. (2016) | 1984–2013 | County | Oct | 50% of the data split randomly | Random forest | 1,130 | 16.7% |
| Kuwata and Shibasaki (2016) | 2008–2013 | County | Oct | 20% of the data split randomly | Deep neural network | 1,142 | 14.0% |
| Jin et al. (2017) | 2001–2015 | County | Aug | 2008-2015 from 6 other states | Ensemble of crop models | 1,286 | 18.6% |
| Crane-Droesch (2018) | 1979–2016 | County | Oct | Out-of-bag samples | Semiparametric neural net | 998 | 13.4% |
| Peng et al. (2018) | 1982–2016 | National | August | Forward CV Out-of-bag samples | Linear regression | 275 | 2.8% |
| Kim et al. (2019) | 2006–2015 | County | Jul–Aug | CV Out-of-bag samples | Deep neural network | 765 | 7.9% |
| Schwalbert et al. (2020) | 2008–2017 | County | Aug | CV Out-of-bag samples | Linear regression | 1,040 | 11.0% |
| Archontoulis et al. (2020) | 2015–2018 | Field | Jun-Aug | Field data | APSIM model | – | 14–20% |

*The proposed model is distinguished with (*).*

TABLE 4 | Summary of state and agricultural district-level model performance.

| ML model | (a) Agricultural district-level forecasts | | | | (b) State-level forecasts | | | |
|---|---|---|---|---|---|---|---|---|
| | RMSE (kg/ha) | RRMSE (%) | MBE (kg/ha) | MDA (%) (2018–2017) | RMSE (kg/ha) | RRMSE (%) | MBE (kg/ha) | MDA (%) (2018–2017) |
| **Linear regression** | 1,556 | 12.99% | 542 | 62.96% | 985 | 8.05% | 305 | 100.00% |
| **LASSO** | 1,364 | 11.39% | 569 | 48.15% | 662 | 5.41% | 321 | 100.00% |
| **XGBoost** | 1,628 | 13.60% | −967 | 40.74% | 1,393 | 11.38% | −1221 | 66.67% |
| **LightGBM** | 1,427 | 11.91% | −607 | 37.04% | 1,087 | 8.88% | −853 | 0.00% |
| **Random forest** | 1,328 | 11.09% | −469 | 29.63% | 945 | 7.72% | −712 | 33.33% |
| **Stacked regression** | 1,266 | 10.57% | −13 | 40.74% | 742 | 6.06% | −251 | 66.67% |
| **Stacked LASSO** | 1,264 | 10.55% | −15 | 40.74% | 632 | 5.17% | −269 | 66.67% |
| **Stacked random f.** | 1,270 | 10.60% | −335 | 29.63% | 836 | 6.83% | −601 | 66.67% |
| **Stacked LightGBM** | 1,242 | 10.37% | −253 | 37.04% | 756 | 6.18% | −503 | 66.67% |
| **Optimized w. ensemble*** | 1,251 | 10.45% | 99 | 33.33% | 608 | 4.97% | −151 | 66.67% |
| **Average ensemble** | 1,262 | 10.54% | −186 | 33.33% | 761 | 6.22% | −432 | 66.67% |
| **EWA** | 1,329 | 11.09% | −468 | 29.63% | 946 | 7.73% | −711 | 33.33% |

*The proposed model is distinguished with (*).*

uncertainty in weather), and thus knowledge of yield can be very valuable. In addition, **Figure 6** proves it further that weighted ensemble models perform better than stacking ensembles even considering all partial weather scenarios.

## Partial Dependence Plots of Optimized Weighted Ensemble

There are extensive studies in the literature (Dietterich, 2000; Shahhosseini et al., 2019a; Shahhosseini et al., 2019b) showing the superiority of more complex machine learning models such as ensemble and neural network models. However, these black-box models lack the interpretability of more simple models and deducing insight from them is more difficult. Friedman (2001) introduced partial dependence plots (PDPs) to explain the dependency of different input features to the predictions made by supervised learning. PDP plots the effect of varying a specific input feature over its marginal distribution on the predicted values.

Let $K$ be a subset of number of input features ($p$), and $K'$ be its complement set, the partial dependence function is defined as follows (Goldstein et al., 2015).

$$\hat{f}_K = E_{x_{K'}}\left[\hat{f}(x_K, x_{K'})\right] = \int \hat{f}(x_K, x_{K'})dP(x_{K'}) \quad (14)$$

in which $dP(x_{K'})$ is the marginal probability distribution of $x_{K'}$. Equation (14) can be estimated as the average of predictions using training data. Let $n$ be the number of training data points, and $x_{K'}^{(i)}$ be the different observed values of $x_{K'}$. Then, the estimation is as follows (Molnar, 2019).

$$\hat{f}_K = \frac{1}{n}\sum \hat{f}\left(x_K, x_{K'}^{(i)}\right) \quad (15)$$

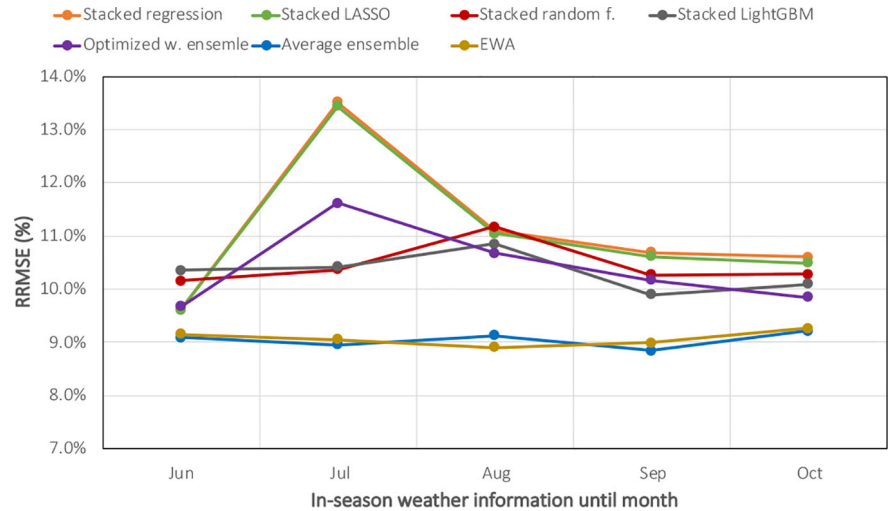

**FIGURE 6** | Evaluating machine learning ensembles when having partial in-season weather knowledge. The X-axis shows the in-season weather information from planting until June, July, August, September, or October.

The proposed optimized weighted ensemble presented in *Analyzed Models* is a weighted average of the base learners' predictions with optimal weights. Therefore, based on equation (15), it can be mathematically proved that the partial dependency estimates of optimized weighted ensemble model for a specific feature is the weighted average of partial dependency estimates of the base learners with same optimal weights. Assuming $\hat{g}_K$ as the partial dependence estimate of optimized weighted average ensemble, $\hat{f}_{ki}$ as partial dependence estimates of base learner $i$ ($i \in [1,m]$), we could write:

$$\hat{g}_K = \frac{1}{n}\sum \hat{g}\left(x_K, x_{K'}^{(i)}\right) =$$
$$\frac{1}{n}\sum \left[w_1\hat{f}_1\left(x_K, x_{K'}^{(i)}\right) + w_2\hat{f}_2\left(x_K, x_{K'}^{(i)}\right) + \ldots + w_m\hat{f}_m\left(x_K, x_{K'}^{(i)}\right)\right] =$$
$$\frac{w_1}{n}\sum \hat{f}_1\left(x_K, x_{K'}^{(i)}\right) + \frac{w_2}{n}\sum \hat{f}_2\left(x_K, x_{K'}^{(i)}\right) + \ldots + \frac{w_m}{n}\sum \hat{f}_m\left(x_K, x_{K'}^{(i)}\right) =$$
$$w_1\hat{f}_{k1} + w_2\hat{f}_{k2} + \ldots + w_m\hat{f}_{km}$$
(16)

Hence, partial dependency plots (PDPs) of input features were prepared after calculating partial dependency estimates of the proposed ensemble model (see **Figure 7**). As the PDPs suggest, increasing some weather features such as water vapor pressure (week 22) and precipitation (weeks 21 and 41) will result in predicting lower corn yields by optimized weighted ensemble model. On the other hand, higher minimum temperature in 19[th] week of the year and higher shortwave radiation (week 29) lead to higher predicted yields. Lastly, earlier planting progress until 19[th] week of the year (higher cumulative planting progress in percentage) will results in lower predictions, while the predictions are almost indifferent to changes in the most influential soil properties. Of interest is the "week" that a feature has a strong impact on yields. The features of constructed model (*e.g.* minimum temperature) are most sensitive in different time periods, and some periods are before the crops are planted. This suggests that conditions before planting are important for accurate yield predictions and justifies our approach of using weather data before planting.

## Feature Importance

Gaining understanding of the data is one of the objectives of building machine learning models. Many models such as decision tree, random forest, and gradient boosting have natural ways of quantifying the importance of input features. However, interpreting the features for more complex models like ensembles and deep neural network models are more difficult, making these models black-box. An approach to estimate the relative influence of each input feature for these black-box models, especially for ensemble models is introduced here. This method is based on partial dependency of input features. Essentially, it can be derived from PDPs that input features that have more variability in their PDP, are more influential in the final predictions made by the ML model (Greenwell et al., 2018). Consequently, the features for which the PDP is flat are likely to be less important than input variables with more variable PDP across range of their values.

To this end, sample standard deviation of the partial dependency values for optimized weighted ensemble calculated earlier is used as a measure of variable importance. In other words, the predictors with higher sample standard deviation are more important features. Assuming $k$ levels for the $i$th input feature and based on $\hat{g}_i(x_{ij})$ calculated earlier in equation (16), we can define importance of features as follows.

$$importance(x_i) = \sqrt{\frac{1}{k-1}\sum_{j=1}^{k}\left[\hat{g}_i(x_{ij}) - \frac{1}{k}\sum_{j=1}^{k}\hat{g}_i(x_{ij})\right]^2} \quad (17)$$

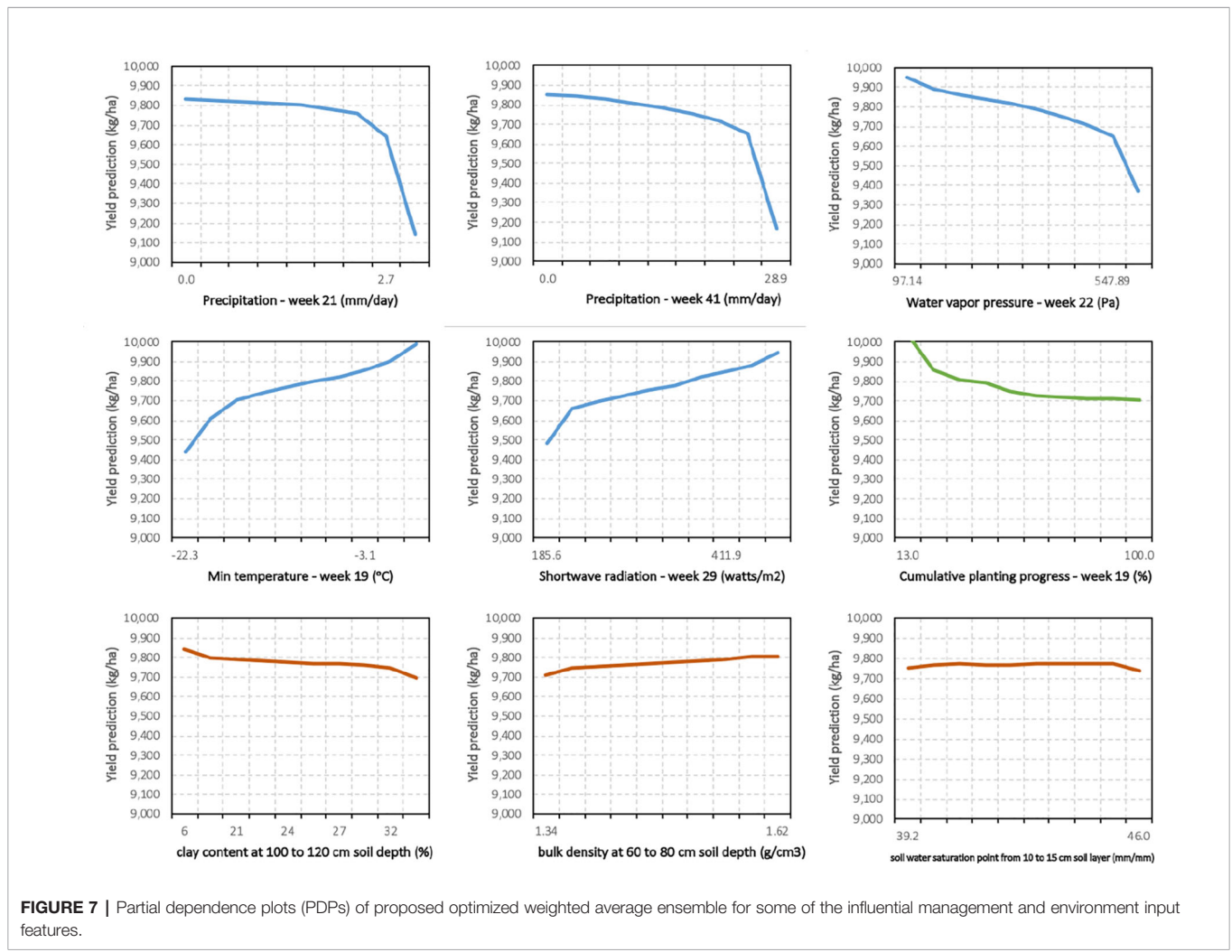

**FIGURE 7** | Partial dependence plots (PDPs) of proposed optimized weighted average ensemble for some of the influential management and environment input features.

**TABLE 5** | Feature importance from optimized weighted ensemble: Top 20 input features.

| | Feature name | Week | Importance |
|---|---|---|---|
| 1 | yield_trend (kg/ha) | – | 1,711.65 |
| 2 | yield_avg (kg/ha) | – | 1,257.70 |
| 3 | precipitation (mm/day) | 21 | 221.14 |
| 4 | precipitation (mm/day) | 41 | 215.32 |
| 5 | water vapor pressure (Pa) | 22 | 164.14 |
| 6 | minimum temperature (°C) | 19 | 155.29 |
| 7 | shortwave radiation (watts/m$^2$) | 29 | 129.36 |
| 8 | water vapor pressure (Pa) | 26 | 120.49 |
| 9 | precipitation (mm/day) | 34 | 115.61 |
| 10 | shortwave radiation (watts/m$^2$) | 44 | 109.33 |
| 11 | water vapor pressure (Pa) | 30 | 108.87 |
| 12 | minimum temperature (°C) | 33 | 107.03 |
| 13 | Cumulative planting progress (%) | 19 | 106.04 |
| 14 | precipitation (mm/day) | 32 | 89.15 |
| 15 | precipitation (mm/day) | 38 | 79.95 |
| 16 | shortwave radiation (watts/m2) | 37 | 77.77 |
| 17 | precipitation (mm/day) | 18 | 76.23 |
| 18 | shortwave radiation (watts/m2) | 27 | 75.73 |
| 19 | minimum temperature (°C) | 28 | 75.07 |
| 20 | shortwave radiation (watts/m2) | 35 | 59.94 |

**Table 5** presents the feature importance results for the top 20 input variables found by optimized weighted ensemble model. Based on the proposed feature importance method, the constructed features for capturing yield's trend, namely yield_trend and yield_avg, are the most important features. All other features from the top 20 input variables consisted of weather parameters along with cumulative planting progress until 19$^{th}$ week of the year. In addition, it seems that weather in weeks 18–24 (May 1$^{st}$ to June 1$^{st}$) is of greater importance compared to weather in other periods of the year.

The framework developed here can be expended to more US states. In addition, more input features such as forecasted weather data, and N-fertilization inputs by county can be added that may result in even higher prediction accuracy. This is something to be explored in the future along with procedures to forecast corn yields with more extensive input features. Further, the developed machines learning models can be used to provide insight into key factors which determine inter-annual yield variability and therefore inform plant breeders and agronomists.

# CONCLUSION

Motivated by the needs to forecast crop yields as early as possible and across scales as well as compare the effectiveness of ensemble learning for ecological problems, especially when there are temporal and spatial correlations in the data, we designed a machine learning based framework to forecast corn yield using weather, soil, plant population, and planting date data.

Several ensemble models were designed using blocked sequential procedure to generate out-of-bag predictions. In addition, an optimized weighted ensemble model was proposed that accounts for both bias and variance of predictions and makes use of out-of-bag predictions to find the optimal weight to combine multiple base learners. The forecasts considered two weather scenarios: complete knowledge of in-season weather and partial knowledge of in-season weather (weather information until June 1$^{st}$, July 1$^{st}$, August 1$^{st}$, September 1$^{st}$, and October 1$^{st}$) and three scales: county, agricultural district, and state levels. The prediction results of the scenario of having partial in-season weather demonstrated that ample corn yield forecasts can be made as early as June 1$^{st}$. Comparing the proposed model with the existing models in the literature, it was demonstrated that the proposed optimized ensemble model is capable of making improved yield forecasts compared to existing ML based models. Furthermore, weighted average ensembles were the leaders among all developed ML models and stacked ensemble models could not perform favorably due to non-IID nature of data set. In addition, a method to find partial dependency and consequently feature importance of optimized weighted ensemble model is proposed which can find the marginal effect of varying each input variable on the ensemble predictions, and rank the input features based on the variability of their partial dependence plots (PDPs). The procedure proposed here for finding partial dependency and feature importance for optimized weighted ensemble model can be easily applied on other ensemble models.

This study is subject to a few limitations, which suggest future research directions. Firstly, it was shown that stacked ensemble models suffer from non-IID nature of the data, and blocked sequential procedure could not help those models predict better than base learners. Working more on the cross-validation procedure to generate improved out-of-bag predictions that emulate test observations better can be considered as a future research direction. Secondly, the performance of ensemble modeling is dependent on the diversity of the selected base ML models, and finding models that are diverse enough is a challenge that needs to be addressed. Therefore, quantifying base models' diversity in order to select more diverse models to create better-performing ensembles can be thought of as future research recommendations. Lastly, adding more input features such as forecasted weather data and N-fertilization inputs by county can improve the model performance. Future research can be done on what additional features should be collected and analysis can be conducted on prediction model.

# DATA AVAILABILITY STATEMENT

The raw data supporting the conclusions of this article will be made available by the authors, without undue reservation, to any qualified researcher.

# AUTHOR CONTRIBUTIONS

MS is the lead author. He conducted the research and wrote the first draft of the manuscript. GH secures funding for this study, oversees the research, reviews and edits the manuscript. SA provides the data and guidance for the research. He also reviews and edits the manuscript.

# ACKNOWLEDGMENTS

This work was partially supported by the National Science Foundation under the LEAP HI and GOALI programs (grant number 1830478) and under the EAGER program (grant number 1842097). This work was also partially supported by the Plant Sciences Institute at Iowa State University.

**Conflict of Interest:** The authors declare that the research was conducted in the absence of any commercial or financial relationships that could be construed as a potential conflict of interest.